# Multimodal transformers with elemental priors for phase classification of X-ray diffraction spectra

Kangyu Ji[1,2]*, Fang Sheng[1], Tianran Liu[1,2], Basita Das[1,2], Tonio Buonassisi[1,2]

[1]Department of Mechanical Engineering, Massachusetts Institute of Technology, 77 Massachusetts Ave, Cambridge, Massachusetts, 02139, United States

[2]Research Laboratory of Electronics, Massachusetts Institute of Technology, 77 Massachusetts Ave, Cambridge, Massachusetts, 02139, United States

*Corresponding to Kangyu Ji (axvcb1597382@gmail.com)

**Abstract**

Classifying a crystalline solid's phase using X-ray diffraction (XRD) is a challenging endeavor, first because this is a poorly constrained problem as there are nearly limitless candidate phases to compare against a given experimental spectrum, and second because experimental signals are confounded by overlapping peaks, preferred orientations, and phase mixtures. To address this challenge, we develop Chem-XRD, a multimodal framework based on vision transformer (ViT) architecture with two defining features: (i) crystallographic data is constrained by elemental priors and (ii) phase is classified according to probabilities (not absolutes). Elemental information can be extracted from pre-synthesis precursors or post-synthesis elemental analysis. By combining structural and elemental modalities, Chem-XRD simultaneously predicts both the number and identity of phases of lead-halide perovskite materials and their decomposition products. Through integrated gradient calculations, we show that the model can adeptly adjust the contributions of structural and elemental modalities toward the final prediction of phase identification, achieving a level of interpretability beyond what self-attention mechanisms can achieve.

## Introduction

X-ray diffraction (XRD)[1] is a common characterization technique used to classify the phases of crystalline solids including metals, semiconductors[2], and pharmaceuticals[3]. Its non-destructive nature enables its broad applications from quality control in manufacturing to materials discovery in self-driving labs[4]. However, for materials systems with unknown structure, similar crystal structures, and/or impurity phases, analyzing XRD data is challenging even for domain experts. By nature, XRD is a poorly constrained problem (the number of candidate compounds to fit a given spectrum is nearly limitless). Furthermore, signal confounders include overlapping peaks, preferred crystal orientations, and phase mixtures. This has led to challenges identifying crystalline phases in recent autonomous-lab campaigns[5,6].

The advancement of machine learning (ML) has opened new possibilities for XRD analysis[7]. Classification of crystal structures[8–10] and extraction of lattice parameters[11] based on convolutional neural networks (CNNs) are promising in speeding up the analysis process. While these methods assume phase purity, most samples in labs contain multiple phases, which can arise from metastable intermediates, phase transformations, and degradation impurities. Recently, multi-phase identification has been achieved in powder XRDs of inorganic oxide chemical space consisting of Sr-Li-Al-O quaternary elements[12]. Models that are robust to preferred crystal orientation have been developed to monitor the *in-situ* thermal phase transformation of $BaTiO_3$ perovskite[13] and identify reaction products during solid-state $Li_7La_3Zr_2O_{12}$ synthesis[14]. However, most ML-based phase-identification methods in literature use only one input — the diffraction pattern itself — and often attempt to identify phase with certainty, mirroring a human scientist. In contrast, modern ML classifiers increasingly rely on multiple input modalities and output a rich range of probabilities, which can reflect uncertainties.

Multimodal learning[15] is a machine learning technique that integrates different modalities (such as images, videos, and texts) to improve model performance. The integration process of different modalities (*i.e.*, data fusion)[16] can occur at the early, intermediate, or late stages of model pipelines. Late data integration, or late fusion, involves the calculation of joint probabilities from the probability of individual modality. Successful late fusion in XRD analysis combines probabilities from 50 ensembled XRD models and elemental composition probability from energy dispersive spectroscopy (EDS) measurement, leading to improved phase prediction accuracy in the Co-Al-Ni chemical space[13]. However, such approaches are not able to learn correlations between features from modalities. To solve this, multimodal models with *early* fusion approaches, such as large language models (LLMs)[17,18], vision transformers (ViTs)[19–21], and CLIP-like models[22–24], are showing promising results that surpass late fusion methods in natural language processing[17], biology[18], and material science[21]. Such models also have the advantage of accommodating input strings of varying lengths, which allow flexibility to vary the spectral range or step size in an XRD measurement, or vary the number of elements in a compound, without constraining input string/matrix length or zero padding.

Here, we developed a multi-modal XRD analysis framework, Chem-XRD, based on the ViT architecture by early fusion of crystallographic XRD data and elemental text priors. The text inputs that contain elemental information can be obtained from either the pre-synthesis chemical precursors or post-synthesis elemental composition analysis, *e.g.*, *via* scanning electron microscope (SEM)-based energy-dispersive X-ray spectroscopy (EDS). In this manuscript, we apply the term "elemental information" to encompass both individual elements (*e.g.*, Cs, Pb, Cl, Br, I…) as well as those forming small molecules that fit within the crystalline lattice (*e.g.*, formamidinium, FA; and methylammonium, MA). Our Chem-XRD model simultaneously predicts the number and identity(ies) of phase(s) generating a given XRD spectrum, often with improved accuracy compared to XRD alone. We demonstrate Chem-XRD for halide perovskite materials[25], as halide perovskites are known for their propensity to form impurity phases and degradation products. We trained the model using 30 million simulated halide perovskite XRD spectra and applied it to resolve 4 experimental XRDs including single-phase and mixed-phase samples, with thin-film and microcrystal morphologies, and with different synthesis recipes and testing conditions. We interpret the model through integrated gradient calculations of Chem-XRD, revealing the decision-making process of our model beyond what self-attention mechanisms can achieve.

## Results

We designed an elementally-informed XRD data collection and curation pipeline based on real-world material synthesis workflows (**Figure 1**). We started the pipeline by listing the elemental precursors and the target materials for synthesis. We then collected data from two primary sources: the Inorganic Crystal Structure Database (ICSD) and relevant publications in material science (**Supplementary Table 1**). We extracted 338 Crystallographic Information Files (CIFs) from the ICSD, which contained structural information of space groups, lattice parameters, and atomic coordinates, as well as synthesis/testing conditions including temperature and pressure. (**Supplementary Table 1**). The down-selected list of CIFs includes 22 compounds from 9 crystallographic space groups and 8 elements. The combination of different elements creates 18 chemical spaces that contain phase mixtures.

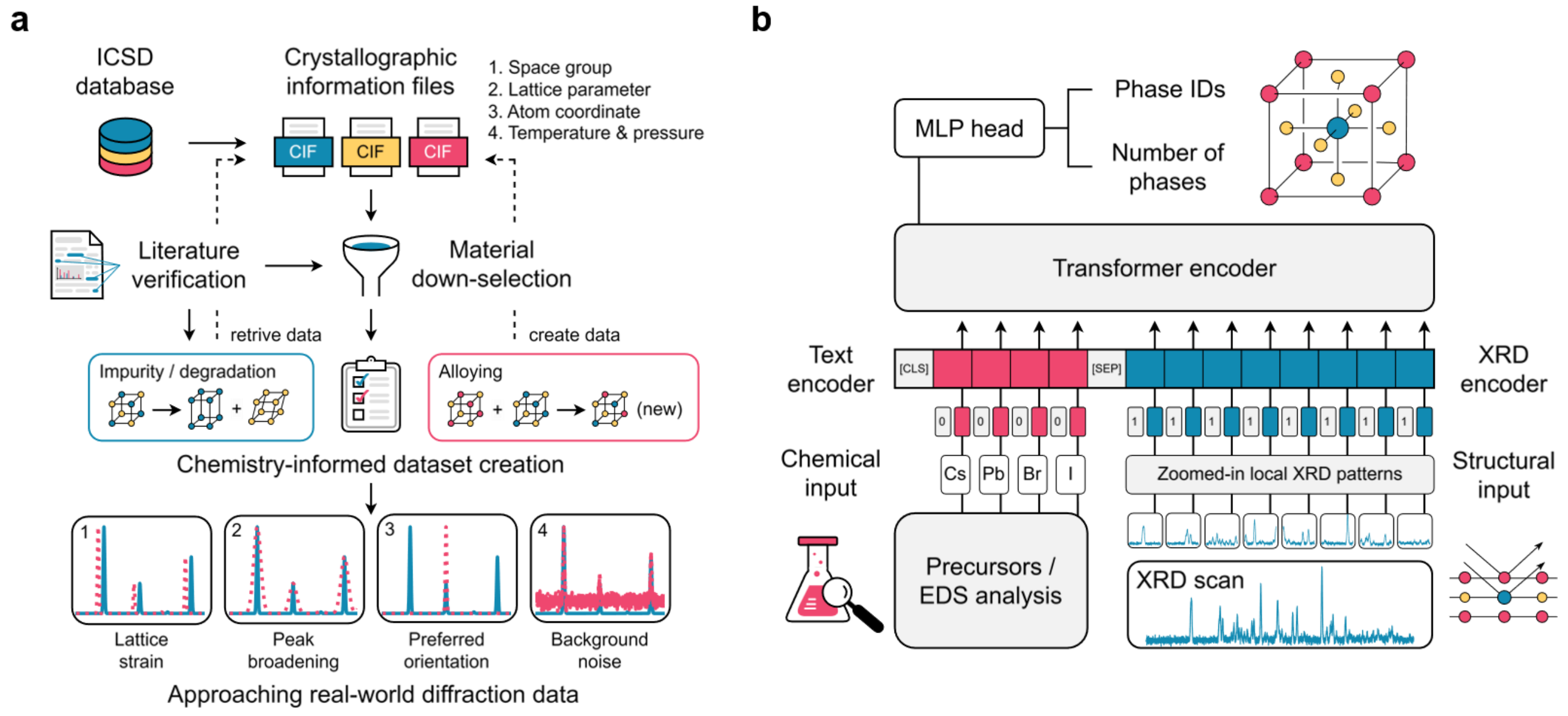

**Fig. 1. Overview of the XRD data collection pipeline and the multimodal Chem-XRD architecture for materials phase identification. a**, Database construction workflow combining ICSD crystallographic data with chemistry-informed dataset creation, including impurity phases, degradation products, and alloyed materials. XRD patterns are simulated by incorporating lattice strain, peak broadening, preferred orientation, and background noise to approximate real-world measurements. **b**, Chem-XRD integrates elemental and structural inputs through the transformer encoder and the multilayer perceptron (MLP) head for phase identification. The modal types are differentiated using token-type embeddings of 0 (elemental) and 1 (structural). Special tokens [CLS] and [SEP] are added to mark the beginning and end of the elemental modality, respectively.

To simulate real-world material synthesis where material degradation, phase transformation, and composition alloying may occur, we expanded the dataset by i) retrieving additional ICSD entries of impurities and degradation products of each chemical space, and ii) generating new structural files through elemental substitution and alloying based on recent publications (**Figure 1a**, also see **Methods**). Lastly, we implemented four data augmentation steps to approach real-world diffraction data characteristics[13,12,26]: i) introduction of lattice strain effects; ii) incorporation of peak broadening phenomena; iii) simulation of preferred orientation effects; and iv) addition of background noise to simulate realistic measurement conditions. The final synthetic dataset containing 66 compounds is visualized through t-distributed stochastic neighbor embedding (t-SNE) in **Supplementary Figures 1** and **2**, where we observed significant overlap between structurally similar phases. These overlaps demonstrate the challenge of phase identification using only the XRD modality.

Our Chem-XRD architecture is shown in **Figure 1b**, developed upon a multimodal vision and language model reported previously[19]. The model accepts elemental (text) and structural (XRD) modalities as inputs. A list of elements, without stoichiometry ratios, obtained either from precursors or energy dispersive spectroscopy (EDS) analysis, is fed into the text encoder to generate text embeddings. We incorporate precursor information as it is readily available during the material synthesis and does not require additional characterization. It can also include organic compounds that EDS cannot detect, useful for the classification of inorganic-organic compounds (such as metal-organic frameworks and hybrid organic-inorganic perovskites). We note that reaction conditions can also be incorporated to extend the current priors, as they can be easily modified in text format. For structural modality, the XRD signal is segmented into zoomed-in local XRD patterns (patches) at fixed sequence lengths, following a similar strategy to the image segmentation method in conventional ViT models. The zoomed-in patches are fed into the XRD encoder to create XRD embeddings. Next, token-type embeddings and positional embeddings are added to both XRD and elemental embeddings to encode modality types and sequential orders. A multilayer perceptron (MLP) head is attached to the transformer output to predict the number of phases and the corresponding material phase IDs.

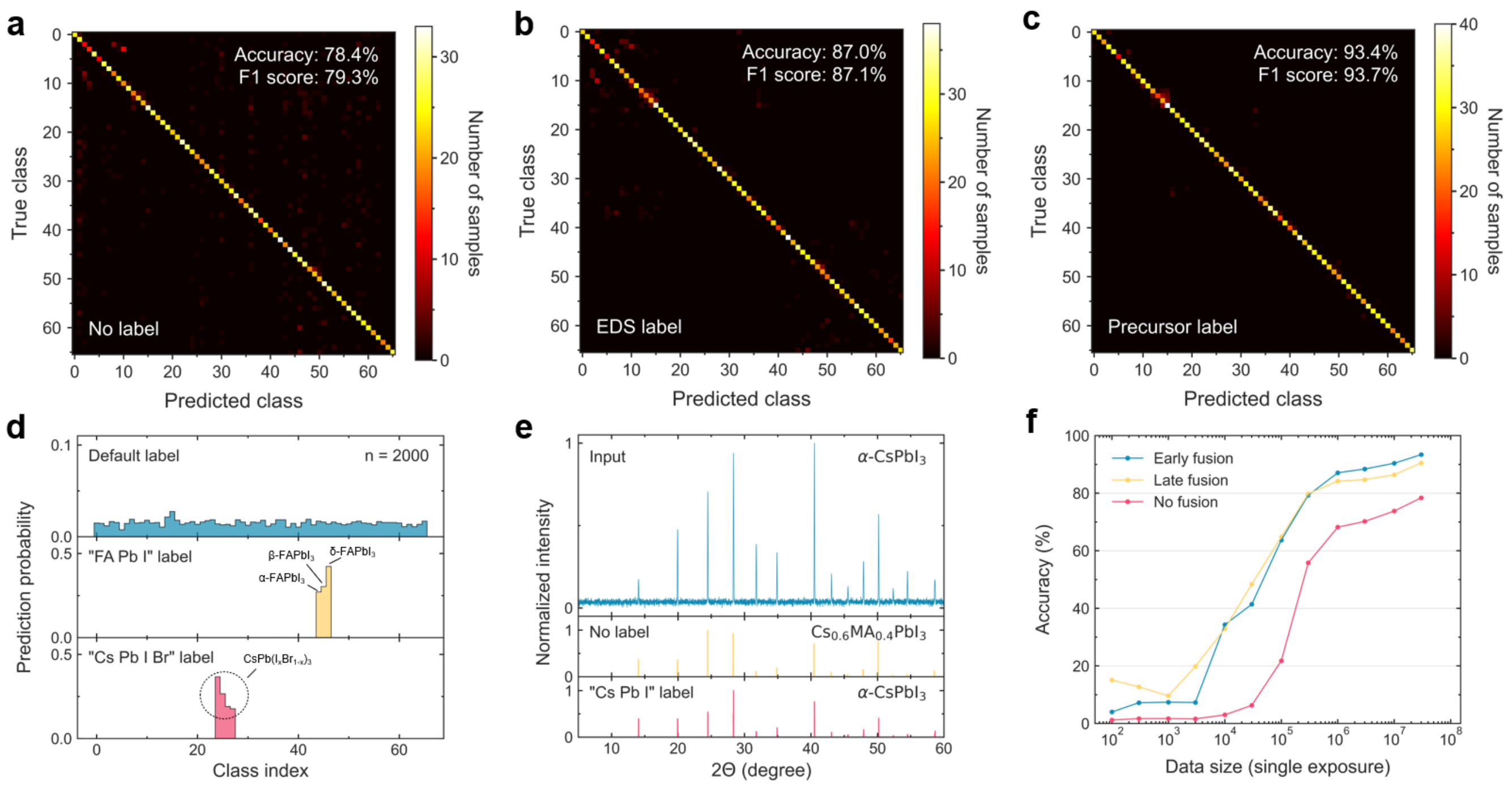

**Fig. 2. Elemental priors enable accurate XRD phase classification in Chem-XRD. a-c**, Confusion matrix of single-phase classification for Chem-XRD models with various levels of elemental priors: (**a**) no elemental label, (**b**) EDS label, and (**c**) precursor label. The accuracy and F1 score were listed on the top right of the matrix. **d**, The probability distribution of model output with various elemental labels. **e**, An example of a wrongly classified phase by models without elemental prior, and the corresponding reference XRD patterns. **f**, The model accuracy against training data shows improved performance with the early fusion strategy (for data sizes >$10^6$). The late fusion model uses the combined probability between the XRD-only model (*i.e.*, no fusion) and the elemental probability. The elemental probability was assigned on a sample-by-sample basis, with each sample within a given chemical space receiving an equal probability. For instance, if a chemical space contains three samples, each sample is assigned a probability of 33%, as stoichiometry information is not utilized. A local XRD window length of 0.5° was used for all models (where the XRD patch size was set to 50). To reduce model overfitting, each datum (spectrum) was used only once.

We first evaluated Chem-XRD in single-phase XRD classification (**Figure 2**). Our baseline model, without incorporating elemental priors, achieved an accuracy of 78.4% and an F1 score of 79.3%. The accuracy improved to 86.9% with element (i.e. EDS) labels and further increased to 93.4% with element and organic precursor labels (**Figure 2a-c**). To study the effect of elemental priors on model prediction, we deliberately replaced the correct precursor labels with arbitrary labels. When using the FA-Pb-I label for all XRD patterns, the model yielded a prediction distribution centering around phase variations of organic-inorganic $FAPbI_3$ perovskite, while the predictions for other compositions were suppressed (**Figure 2d**). Similarly, using the Cs-Pb-I-Br label confined the model prediction within the mixed halide perovskite family $CsPb(I_xBr_{1-x})_3$ ($0 < x < 1$). These results demonstrate that elemental priors effectively constrain the model prediction to the target chemical space. This allows the model to differentiate structurally similar samples with composition differences. A typical example is shown in **Figure 2e**, where the XRD pattern for α-$CsPbI_3$ is mislabeled as $Cs_{0.6}MA_{0.4}PbI_3$ by the baseline model due to their similar crystal structures and peak positions. With the precursor information, such misclassification can be effectively prevented. We also observe models with elemental priors show faster convergence compared to the baseline, which indicates that the incorporation of elemental modality improves the training efficiency of models (**Supplementary Figure 3**).

We then compared the model performance across different data fusion strategies (**Figure 2f**). We observed the advantage of late fusion over other methods when the training data was low ($< 10^4$), as the joint probability was able to directly exclude candidates outside of the target chemical space — a common approach in the manual XRD analysis workflow. However, as the data size increases, the performance of the early fusion model surpasses that of the late fusion model (see also **Supplementary Table 2** for CNN models). This indicates that early fusion, which combines XRD and elemental features at the input level, can leverage cross-modality correlations to achieve better scalability and accuracy for larger datasets.

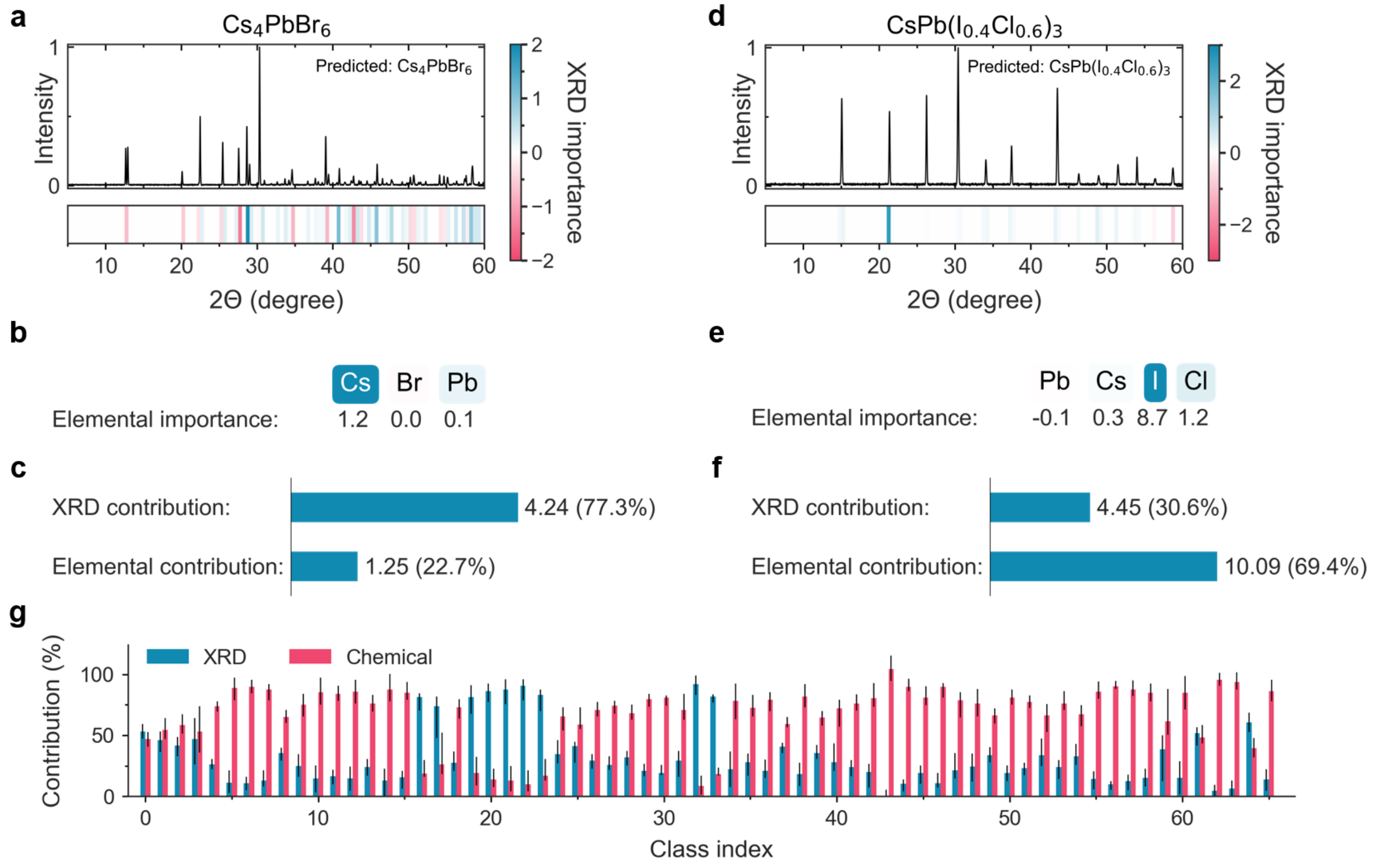

**Fig. 3. Model interpretation of elemental and structural contributions for phase identification. a-c**, Individual contributions of (**a**) XRD and (**b**) elemental modalities, as well as (**c**) the overall contributions to the final ranking score for classifying $Cs_4PbBr_6$. **d–e**, Contributions for classifying $CsPb(I_{0.4}Cl_{0.6})_3$. **g**, Distribution of XRD and elemental contributions in percentage across different material classes in the test set, showing the model adaptively weights structural and compositional information depending on the specific phase system. Error bars show the lower and upper quartiles of the distribution. The importance score for each modality was obtained by approximating the integral of gradients of model output back to the inputs of the corresponding modality along the network path, with background calibration from a baseline (*i.e.*, with empty inputs)[27] (see also Methods).

To interpret the model decisions, we first examined the cross-correlation between elemental and XRD modalities through the self-attention layers of the last transformer block (**Supplementary Figure 4**). However, the cross-attention scores from elemental priors to XRD patterns are not easy to understand, as there is a lack of scientific rationale between individual elements and structural patterns.

To address this, we adopted an alternative approach inspired by the multimodal interpretation of visual-question-answering models, where we calculated the integrated gradients[27] from the model outputs back to the inputs (**Figure 3**, also see **Methods**). For the Cs-Pb-Br chemical space, where several possible phases can form with the same elemental precursors, we found the model relied on XRD inputs over elemental inputs (**Figure 3a–c**). This is reflected in both the low elemental importance scores of individual elements (Cs: 1.1, Pb: 0.1, Br: 0.0) and the dominant XRD contribution (78.3%) over elemental contribution (21.7%). In contrast, for the mixed-halide $CsPb(I_{0.4}Cl_{0.6})_3$, the elemental priors contributed to 73.4% of the total prediction score (**Figure 3d–f**). For simple elemental spaces with only one possible phase, the model primarily relied on elemental priors, where elemental contribution exceeded 90% of the total contribution (**Supplementary Figure 5**). Such adaptive behavior across different material phases is visualized in the contribution plot (**Figure 3g**).

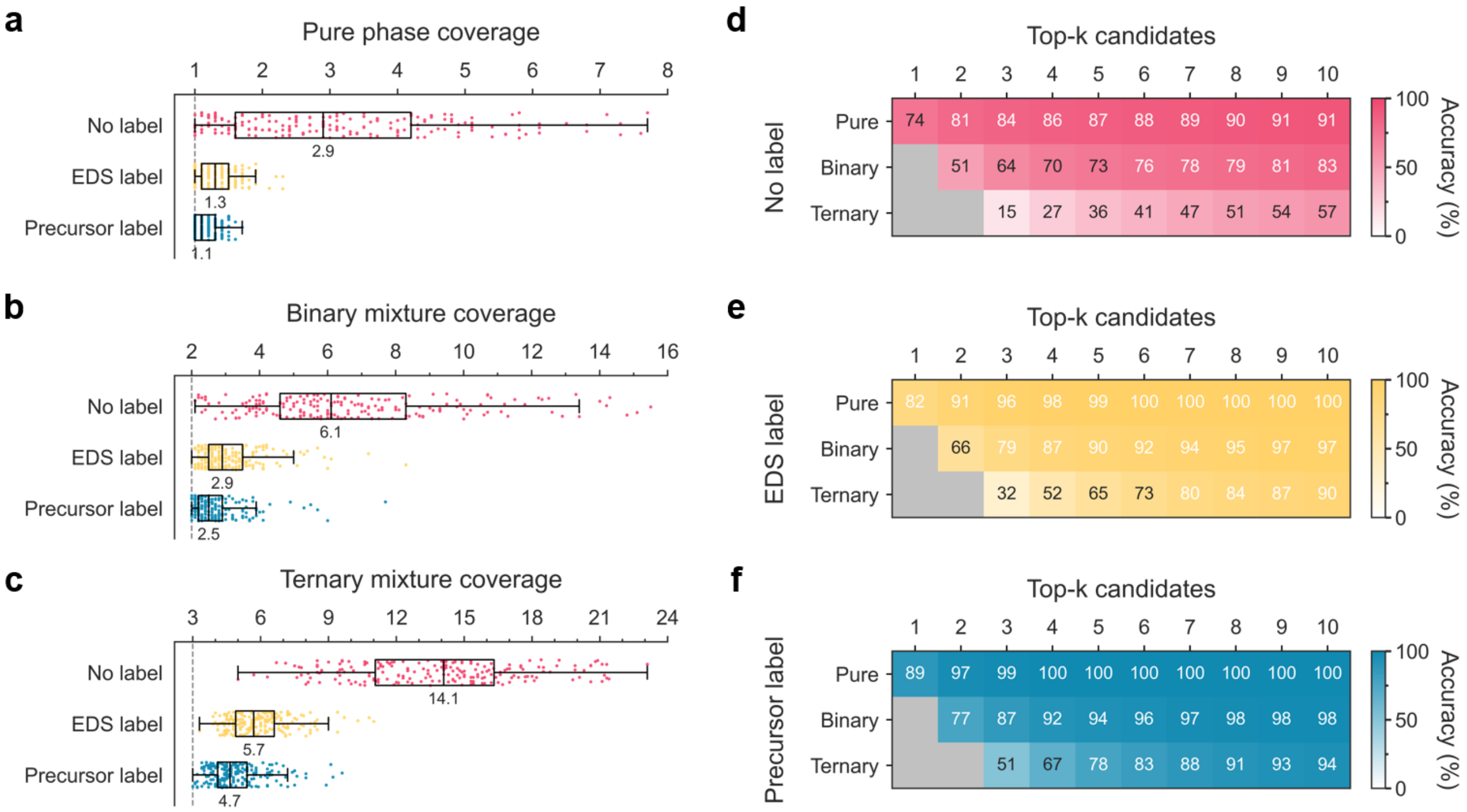

**Fig. 4. Multiphase XRD identification with Chem-XRD. a-c**, Multilabel coverage error distribution showing increasing prediction complexity with (**a**) pure, (**b**) binary, and (**c**) ternary mixtures. Multilabel coverage error statistics were collected

over 2000 simulated samples, with each value averaged from 10 consecutive predictions. **d–f**, Top-*k* accuracy with different elemental priors: (**d**) no label, (**e**) EDS label, and (**f**) precursor label.

We evaluated the performance of Chem-XRD in multiphase identification (**Figure 4**). We trained a single model to predict both the number of phases and the phase IDs with different elemental priors. For the number of phase prediction, the model with precursor inputs achieved 98.9% accuracy for classifying pure phases, 89.2% for binary, and 43.9% for ternary mixtures (**Supplementary Figure 6**). We note the difficulty in identifying ternary mixtures, as we only allow phase mixtures within a given chemical space (*i.e.*, a given elemental set without stoichiometry information), which contains phases with overlapping diffraction peaks and similar phase structures (**Figure 1a**, also see Methods).

We then quantified the model performance in multiphase identification using multilabel coverage error[28] and multilabel top-*k* accuracy scores (see **Methods**). The coverage error was calculated by the minimum number of predictions (sorted from high to low score) required to cover all ground truth phases. We found that precursor priors reduced the coverage error from 2.9 to 1.1 in pure phases, 6.1 to 2.5 in binary, and 14.1 to 4.7 in ternary mixtures (**Figure 4a-c**). This improvement was further demonstrated by the top-*k* accuracy analysis (**Figure 4d-f**). With precursor labels, the model achieved 77% accuracy for binary mixtures in top-2 prediction and 94% in top-5 (**Figure 4f**). For ternary mixtures, predictions with precursor labels achieved 51% accuracy in top-3 and 78% in top-5, outperforming the performance with no elemental priors (65% in top-5) and EDS labels (36% in top-5). We observed similar improvements in multiphase identification when incorporating the elemental modality into the CNN models (**Supplementary Figure 6**).

We used Chem-XRD to predict both the number and nature of phase(s) in experimental lead-halide samples produced from different precursors and synthesis methods, from which XRD spectra were obtained with different measurement protocols (**Figure 5**). We created two aqueous synthesis recipes for $CsPbBr_3$ microcrystals, one with balanced stoichiometry and the other with off-balance stoichiometry containing excess $Cs^+$ ions in the precursor solution. The model correctly identified the target orthorhombic-$CsPbBr_3$ phase (**Figure 5a**) and detected the $Cs_4PbBr_6$ impurity phase in the sample created from Cs-rich precursors (**Figure 5b**). Our model maintained accurate predictions despite using different XRD scanning ranges (5–45° and 10–60°) from the training dataset (5–60°). We then evaluated thin-film perovskite samples, where XRD patterns exhibited strongly preferred orientations. For the spin-coated $MAPbI_3$ thin film, the model identified both α-$MAPbI_3$ and β-$MAPbI_3$ phases (**Figure 5c**), consistent with the Rietveld refinement analysis (**Supplementary Figure 8**). We also prepared a degraded $FAPbI_3$ film, where the α-$FAPbI_3$ underwent phase transformation into the δ-$FAPbI_3$ phase, and decomposed into FAI (which escapes into air) and $PbI_2$ (which remains in the film)[29]. The model successfully identified all of the phases in the top-4 candidates (**Figure 5d**). This task is particularly challenging due to the strong preferred orientation of degradation products, where only one or a few diffraction peaks can be observed. These results demonstrate that Chem-XRD can be used across different chemical spaces and experimental conditions, representative of real-world variances during laboratory research.

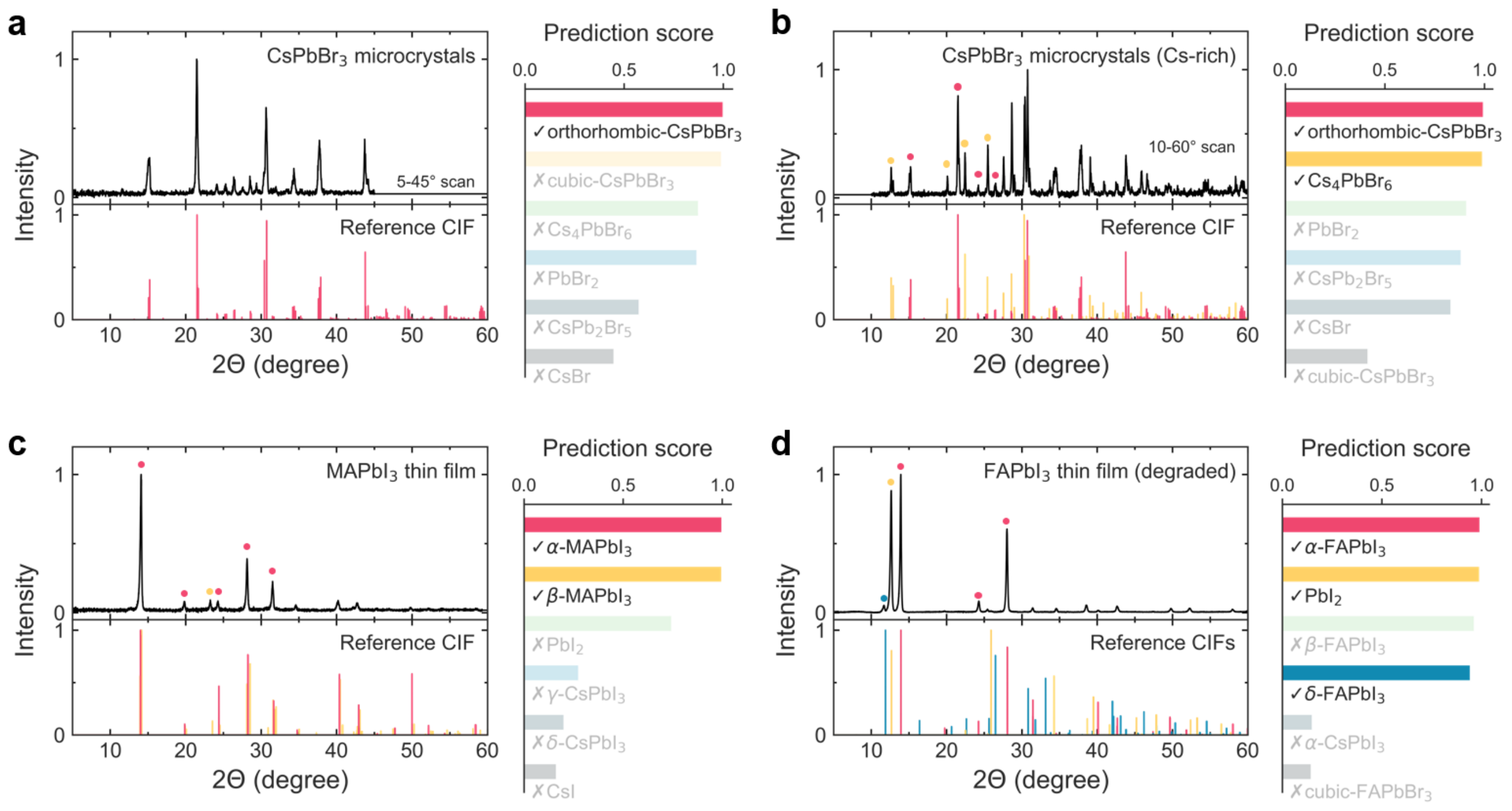

**Fig. 5. Experimental validation with Chem-XRD across different perovskite material systems.** Experimental XRD patterns are plotted against the simulated XRD data from CIF files, with the model prediction scores ranked for top 6 possible candidates. **a**, Phase-pure $CsPbBr_3$ microcrystals. **b**, $CsPbBr_3$ microcrystals with an off-balanced Cs-rich stoichiometry in precursor solutions, leading to the formation of $Cs_4PbBr_6$ impurity. **c**, $MAPbI_3$ thin film on glass. **d**, Degraded $FAPbI_3$ thin film fabricated using non-stoichiometric, $PbI_2$-rich precursors.

## Discussion

We present Chem-XRD, a multimodal transformer for multiphase XRD analysis that leverages elemental information from either precursors or elemental analysis. We develop a chemistry-informed dataset collection pipeline that incorporates literature-verified material synthesis scenarios with impurities, degradation products, and compositional alloying. Models with elemental priors demonstrate improved prediction accuracy and learning efficiency. By adaptively weighting the structural and elemental information, the model can perform phase identification across 18 different chemical spaces. We validate the model using experimental XRD data across different lead-halide materials systems, synthesis conditions, and slightly varying measurement protocols. The model demonstrates good performance in identifying multi-phase samples with strong preferred orientations and peak overlaps.

Our method still faces several challenges. While our model shows promising results on synthetic datasets, validation of experimental XRD patterns is time-consuming, and a complete analysis of one multiphase thin film sample could take hours of effort with domain experts (especially for thin film samples). The model relies on the assumption that the training dataset represents experimental data. However, materials measured in the lab may exhibit variations in lattice parameters compared to their reported values. This discrepancy becomes pronounced when the model is trained on simulated XRD patterns from density functional theory (DFT) calculations rather than experimentally reported values.

Conventional train-test splits method fail for material ID classification because each class corresponds to a unique diffraction pattern rather than a broader category with multiple examples (such as in space group or material dimensionality classification), preventing the model from learning generalizable features. Thus, our model cannot extend its predictions beyond the materials it has been explicitly trained on, making it not suitable for identifying new materials. The effectiveness of precursor labels can also be influenced by unexpected impurities or contaminants that are not included in the elemental priors.

Chem-XRD requires more training to optimize its weights compared to simpler models such as CNNs. However, the transformer model architecture can accommodate inputs of various dimensions and is expected to scale better as datasets grow. This is especially useful when XRD spectra have varying 2θ ranges, and/or the number of elements varies from sample to sample. On the other hand, choosing the right multimodal inputs is crucial for achieving better results than single modality inputs. The elemental priors (text inputs) can include material synthesis recipes with detailed fabrication steps and reaction conditions. The addition of other kinds of modalities can also provide complementary information to improve the phase identification accuracy. In addition, localized optical properties measured through spatially resolved spectroscopy can serve as indirect indicators of material degradation, and electron diffraction patterns from high-resolution transmission electron microscopy (HRTEM) can identify the structure of trace impurities not observable in standard XRD measurements.

While integrated gradient methods offer cross-modality model interpretation, they assume that features contribute independently to predictions. However, in XRD patterns, peak intensities are often correlated (except in highly oriented samples). As a result, a feature may appear to have a negative or negligible effect when, in reality, its contribution is conditional on another feature (**Figure 3**). Future work could focus on detecting and isolating these feature interactions[30] both within and across modalities, which will further improve interpretability.

Finally, collecting a large and accurately labeled multimodal dataset remains a challenge in materials science. Given the improvement of early-fusion performance with datasets greater than $10^6$, it does suggest that larger datasets will be instrumental in realizing a generalizable multi-phase XRD model — either *via* augmented synthetic and/or expert-labeled experimental datasets. Developing a self-driven multimodal characterization platform could be one such path toward creating this dataset.

## Methods

### Dataset preparation

We conducted a literature review and screened material entries from the ICSD based on specific halide perovskite compositions and degradation products. For each material composition, we only collected unique entries for possible space groups by validating the corresponding literature and related papers that cited these ICSD collection codes in XRD data analysis. This systematic data gathering and screening approach enabled us to create a comprehensive lead halide perovskite dataset (**Supplementary Table 1**).

We generated new alloys by loading CIF files for two candidate perovskite structures from the same space group and extracting their lattice parameters and atomic compositions via Python Materials Genomics (pymatgen)[31], an open-source materials analysis library. We listed possible combinations of atomic species (such as mixed organic, mixed metal, or mixed halide), and created new CIF files with composition gradients with a step of 20% by applying proportional strains on baseline structures.

During preprocessing, we applied lattice strains (ranging from -0.2% to 0.2%) in three directions on all CIF files to mimic realistic lattice distortions. We then computed the XRD patterns within a specified 2θ range (5° to 60°) and created an XRD dataset containing the material class IDs, lists of XRD patterns with various lattice strains, the corresponding material compositions, and common element names.

Our dataloader augments the XRD patterns by applying the following steps to the dataset: i) random strain pattern, ii) random individual peak intensities varying from 20-100% of the original pattern, iii) random preferred orientation effects (20% probability) applied to one of the top three highest intensity peaks, iv) random peak broadening based on the Scherrer equation with crystallite sizes ranging from 5 to 20 nm, and v) random background Gaussian noise with standard deviation ranging from $10^{-4}$ to $10^{-1}$ in logspace, where XRD intensities are normalized to [0,1].

**Model architecture**

We developed Chem-XRD from the multimodal vision and language model VisualBert[19]. The model has 12 million parameters, 12 transformer layers, 1 attention head, 50 visual embeddings per XRD token, and 110 XRD embedding token lengths. Each XRD token consists of 50 data points from a local 2θ range of 0.5° with a step size of 0.01°. The natural language tokenizer for the elemental prior was adopted from the BERT model, bert-base-uncased[32], and was extended to include elements and organic molecules that would otherwise be segmented into two separate tokens. The number of class labels was set to the total number of phase IDs in the dataset. For multiphase applications, three additional labels were included to represent the pure, binary, and ternary phase combinations.

**Model training and testing**

The Chem-XRD model is compatible with any consumer-grade GPU that has over 6 GB of memory with a batch size of 20. In this study, we trained the model on an NVIDIA GeForce RTX 4090 GPU for over two days. During training, the dataloader generated new XRD patterns along with the corresponding elemental labels for each training sample. A total of 30 million unique samples were used, resulting in 1.5 million training steps under a batch size of 20. The ADAM optimizer was used with an initial learning rate of $10^{-4}$, which was later reduced to $10^{-5}$ after 10,000 samples. The common cross-entropy loss (CrossEntropyLoss) was used for single-phase classification, and the binary cross-entropy loss with an additional Sigmoid layer (BCEWithLogitsLoss) was used for multi-label classification without any further modification to class weights.

For testing, we used simulated datasets containing single, binary, and ternary mixtures, each containing 2,000 samples. This is equivalent to requesting the dataloader to generate a training batch with a batch size of 2,000. We note the difference (in orders of magnitude) between training and testing data sizes. Evaluation with common k-fold validation approaches with more than millions of samples requires substantial data storage and model run time. We argue a small subset of the randomly generated samples could still be sufficient to evaluate the overall performance of the model. As all the data generated is used only once without repetition during training, the model never directly encounters the testing data. Because of the small number of materials classes evaluated in this study (N = 18), it was not feasible to train the model using best practices of removing an entire class each time an augmented spectrum was used. Consequently, for this study, there is likely information leakage between train and test sets.

## Multilabel evaluation

For a multilabel dataset $(X_i, Y_i)$, where $i = 1 \dots n$, the labels of a sample input $X_i$ are $Y_i \subseteq L$, a subset of all possible labels $L = \{\lambda_j : j = 1 \dots m\}$. The *multilabel coverage error*[28] measures how far we need to go down the ranked list of labels to cover all true labels of a sample,

$$Coverage\ error\ of\ a\ sample\ = \frac{1}{m}\sum_{i=1}^{m} \max_{\lambda \in Y_i} r_i(\lambda)$$

where the rank predicted by the model (from the highest score to lowest) for a label $\lambda$ is denoted as $r_i(\lambda)$. The best coverage error of a dataset is equal to the average number of labels per sample.

The multilabel top-*k* accuracy[28] measures the fraction of samples where all true labels are contained within the top-*k* predicted labels of the model,

$$Top\text{-}k\ accuracy = \frac{1}{n}\sum_{i=1}^{n} I(Top_k(Z_i) = Y_i)$$

where $I(\text{true}) = 1$, $I(\text{false}) = 0$, and $Top_k(Z_i)$ is the top-*k* predicted labels (sorted from highest scores to the lowest scores) of the model prediction $Z_i$ for the sample $X_i$.

## Multimodal interpretation

We implemented the integrated gradient method using Captum[33], an open-source model interpretability library for PyTorch, to extract multimodal feature importances from our Chem-XRD model. The attribution scores for each modality were calculated by integrating the gradients along a straight path from a baseline instance to the input instance[27]. The baseline was defined by setting XRD intensities and text token values to zero (while preserving [CLS] and [SEP] tokens) while maintaining the same sequence length, token type embedding, and positional embedding as the input.

**Sample preparation**

*Materials*

N, N-Dimethylformamide (DMF, anhydrous, 99.8%), dimethyl sulfoxide (DMSO, anhydrous, 99.7+%), diethyl ether (anhydrous, 99.5%), and chlorobenzene (CB, anhydrous, 99.8%) were purchased from Fisher Scientific. Methylammonium iodide (MAI, >99.99%), methylammonium chloride (MACl, >99.99%), and formamidinium iodide (FAI, >99.99%) were purchased from Greatcell Solar. Cesium acetate (CsAc, 99.9%) and hydrobromic acid (HBr, 48%) were purchased from Sigma Aldrich. Lead bromide ($PbBr_2$, >98.0%) and lead iodide ($PbI_2$, 99.99%) were purchased from TCI America. All chemicals were used without further purification.

*Thin film deposition*

The 1.0 M $MAPbI_3$ solution was prepared by dissolving MAI and $PbI_2$ in a mixed solvent of DMF and DMSO (4:1 v/v). The $MAPbI_3$ thin films were deposited via spin-coating at 1000 rpm for 10 s and 5000 rpm for 30 s. 20 s before the end, 150 μL of CB as the anti-solvent was dropped onto the substrate. The films were then thermally annealed at 100 °C for 30 min to remove residual solvent. The $FAPbI_3$ perovskite solution was prepared by dissolving 1.53 M $PbI_2$, 1.4 M FAI, and 0.44 M MACl in a mixed solvent of DMF and DMSO (8:1 v/v). The $FAPbI_3$ perovskite thin films were deposited via spin-coating at 1000 rpm for 10 s, and 5000 rpm for 30 s. 20 s before the end, 700 μL of diethyl ether as the anti-solvent was dropped onto the substrate. Afterward, the films were thermally annealed at 150 °C for 15 min. All processes above are done inside a nitrogen glovebox at 25°C.

*Microcrystal synthesis*

The $CsPbBr_3$ microcrystals were prepared by adding the pre-dissolved 1.0 M $PbBr_2$ into 1.0 M CsAc in aqueous HBr with a given volume ratio at room temperature. The resulting solution was vigorously stirred for 1 min and dried at 130°C with the container cap left open overnight. The baseline sample was synthesized with a $Pb^{2+}$:$Cs^{+}$ stoichiometry ratio of 1:1, and the Cs-rich sample with a ratio of 4:1.

**Experimental X-ray diffraction**

The X-ray diffraction data was collected by a Rigaku SmartLab X-ray diffractometer with Cu-Kα radiation. Measurements were performed in ambient air at room temperature. The background level was subtracted from the raw XRD data with polynomial fits. The XRD signals were calibrated with estimated sample height displacement from the baseline and normalized to [0,1] before inputting into the model.

**Data availability**

The CIF files used for constructing the synthetic datasets are available via the ICSD database (https://icsd.products.fiz-karlsruhe.de/). The experimental XRD data can be downloaded at https://github.com/PV-Lab/chem-xrd.

**Code availability**

All code used in this work was implemented in Python using PyTorch and Matplotlib, and the source code is available at https://github.com/PV-Lab/chem-xrd.

## Acknowledgments

We thank Jordan Cox for XRD advice. We thank Gerd Ceder (UC Berkeley) for discussion with TB concerning multi-modal uncertainty-aware XRD classification. We thank First Solar for support (KJ, FS, BD) and fruitful discussions. This material is partially based upon work supported by the U.S. Department of Energy's Office of Energy Efficiency and Renewable Energy (EERE) under the Solar Energy Technologies Office Award Number DE-EE0010503 (TL).

## Contributions

K.J. conceived and designed the study, after problem definition by T.B. K.J., F.S., and T.L. collected and curated the dataset. K.J. developed and evaluated the Chem-XRD model. T.L., F.S., B.D., and K.J. fabricated the samples. F.S. collected the experimental XRD data. K.J. analyzed and visualized the results. T.B. supervised the project. K.J. wrote and all authors contributed to the edits of the manuscript.

## Competing interests

While some of the authors have patents in perovskite materials and solar photovoltaics, all of the contents of this manuscript are open sourced, and the authors declare that they have no competing interests.

## Table of Contents

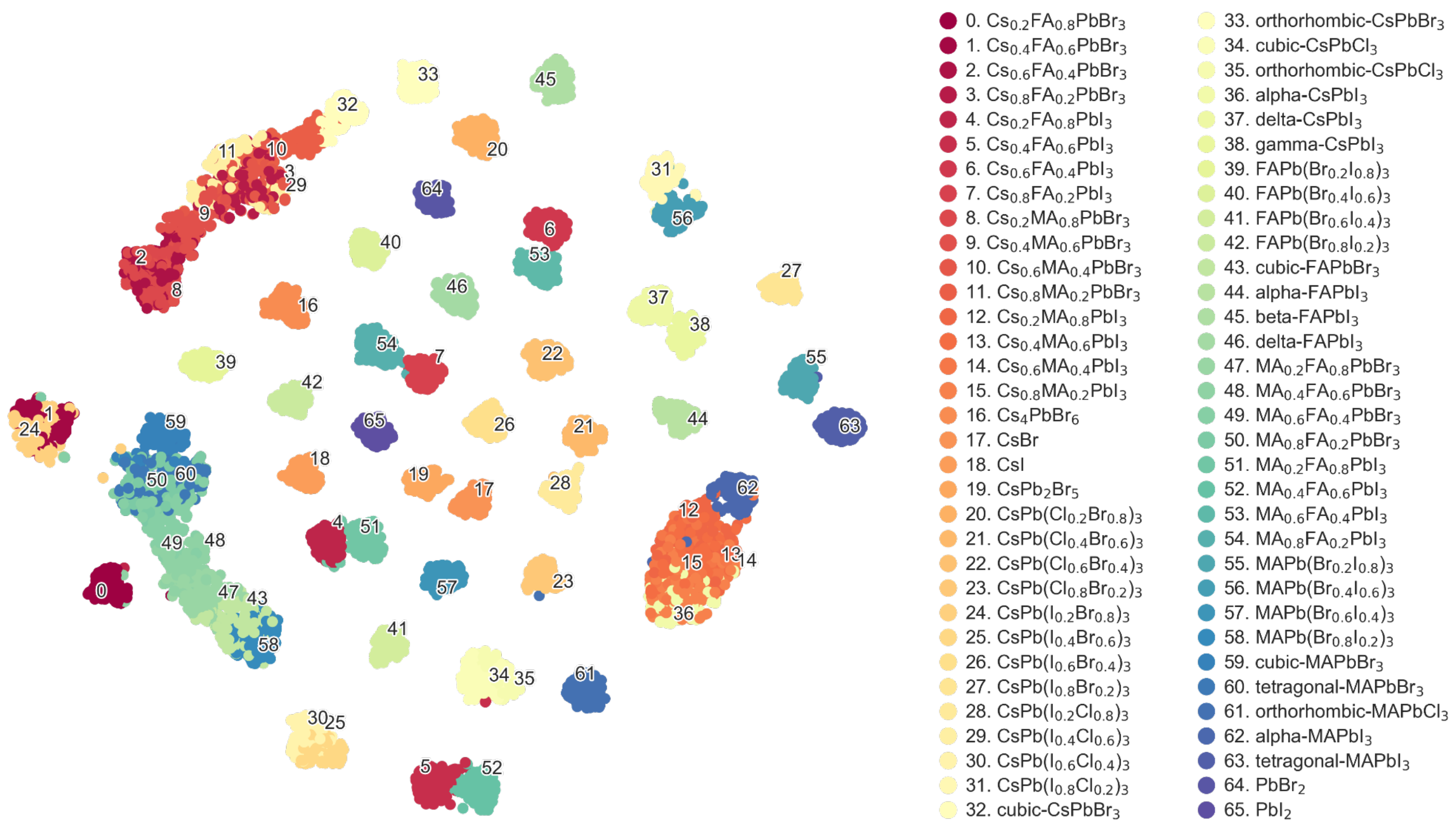

**Supplementary Figure 1. The t-distributed stochastic neighbor embedding (t-SNE) plot of the synthetic XRD dataset without preferred crystal orientations.** Cluster overlaps were observed among various material phases. 10,000 XRD patterns were used to visualize the similarity of XRD inputs in a 2D space. The likelihood of generating highly preferred orientation XRD patterns was set to 0%.

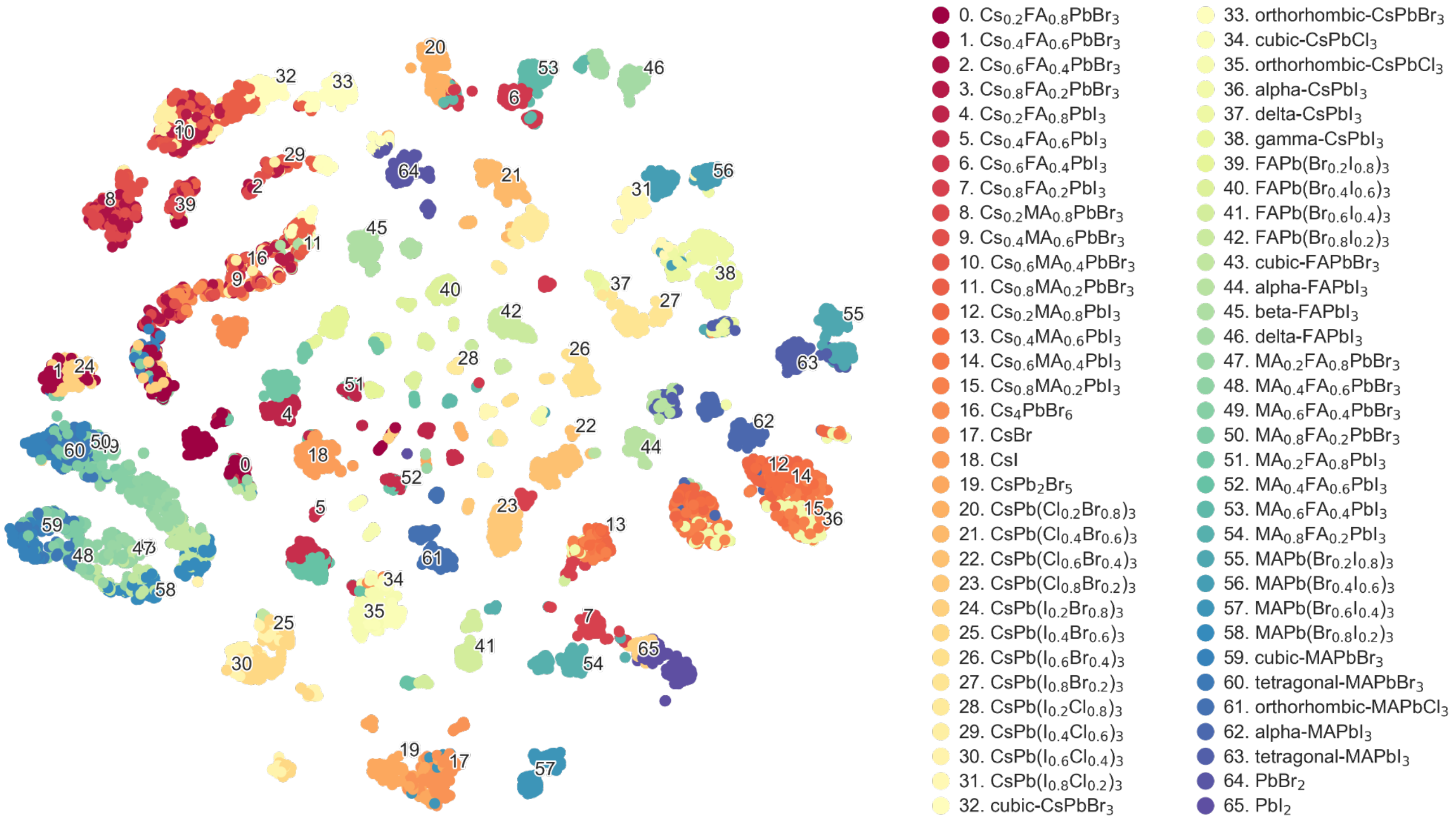

**Supplementary Figure 2. The t-SNE plot of the synthetic XRD dataset with preferred crystal orientations**. The overlapping regions of different material phases are larger compared to the dataset without preferred crystal orientations. 10,000 XRD patterns were used to visualize the similarity of XRD inputs in a 2D space. The likelihood of generating highly preferred orientation XRD patterns was set to 50%.

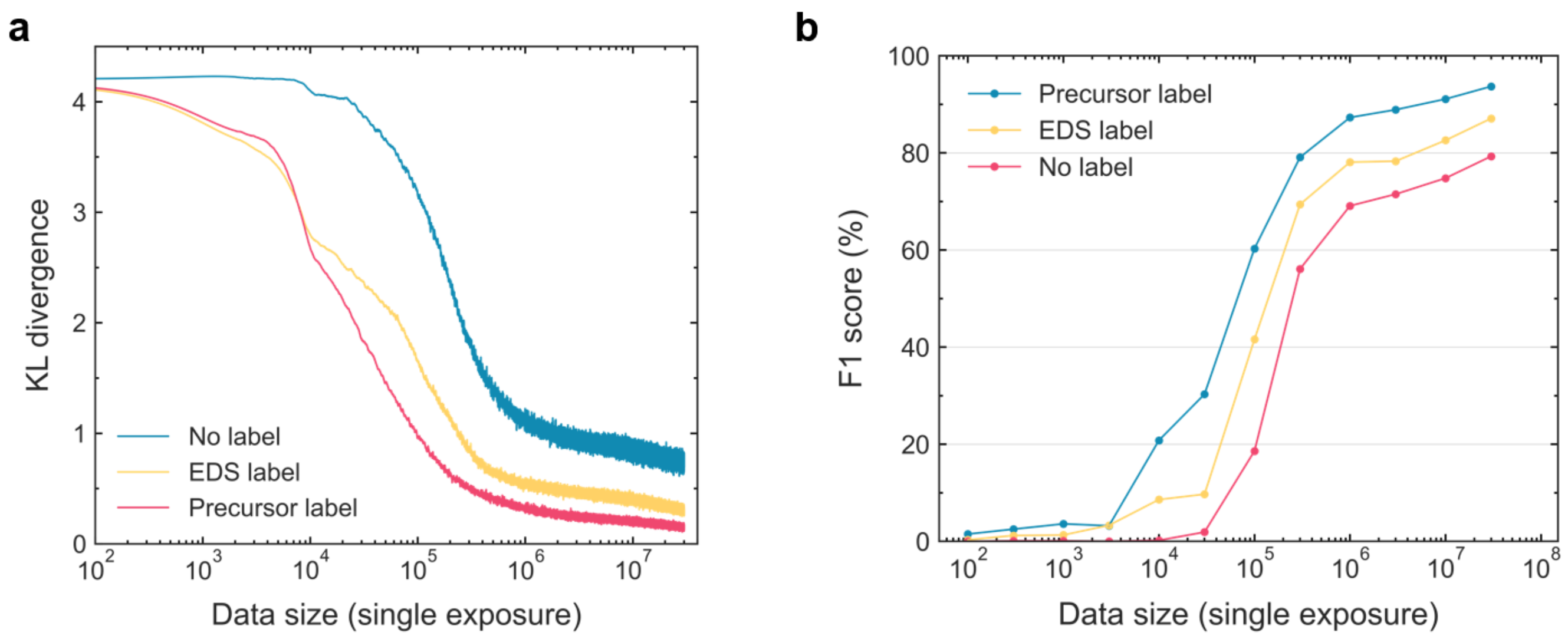

**Supplementary Figure 3. (a) Kullback–Leibler (KL) divergence and (b) F1 score with different elemental priors against the training data size.** The results are plotted against the amount of data used for training. The patch size of the Chem-XRD models was set to 50, corresponding to a local 2θ range of 0.5°. Each model was trained only on a given prompt type. To reduce model overfitting, each datum was used only once. A Savitzky-Golay filter was applied to smooth the KL curves, using a window length of 201 and polynomial order of 3.

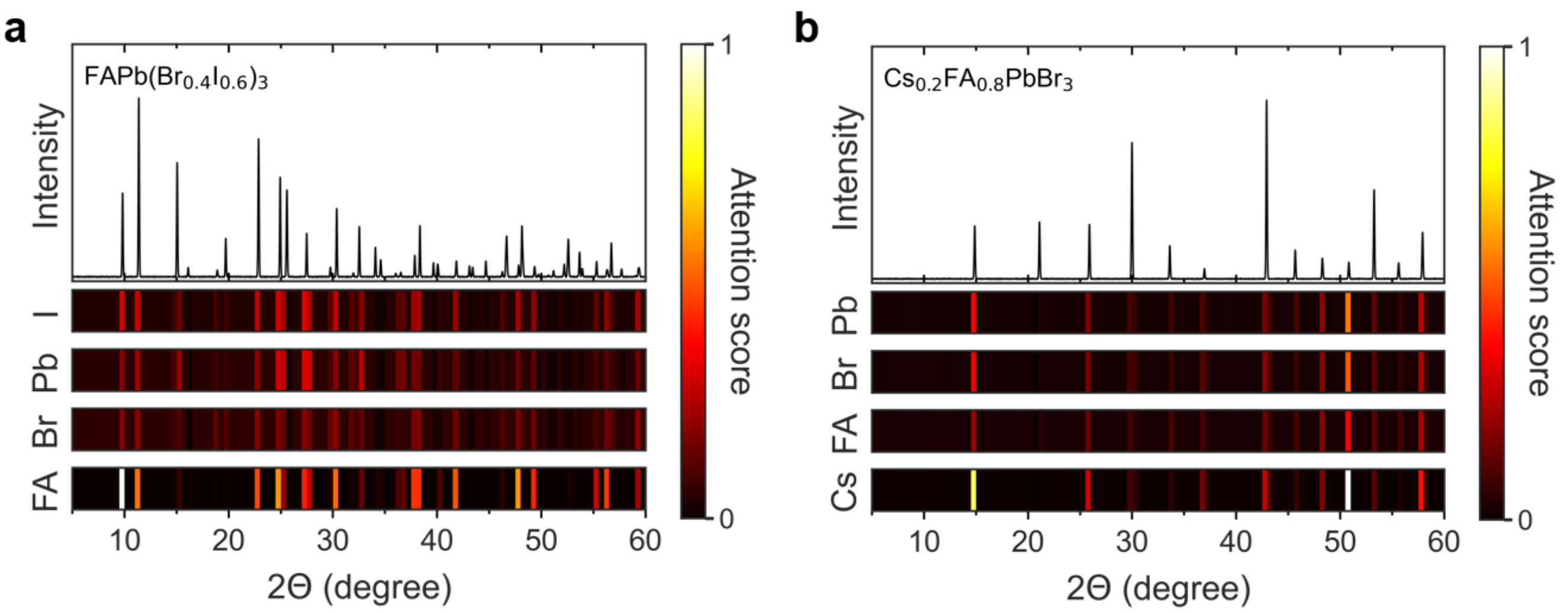

**Supplementary Figure 4. Visualization of cross-attention between modalities in Chem-XRD. a** and **b**, Examples of elemental-to-XRD attention for (**a**) $FAPb(Br_{0.4}I_{0.6})_3$, and (**b**) $Cs_{0.2}FA_{0.8}PbBr_3$. The self-attention scores are extracted from the last layer of the transformer model.

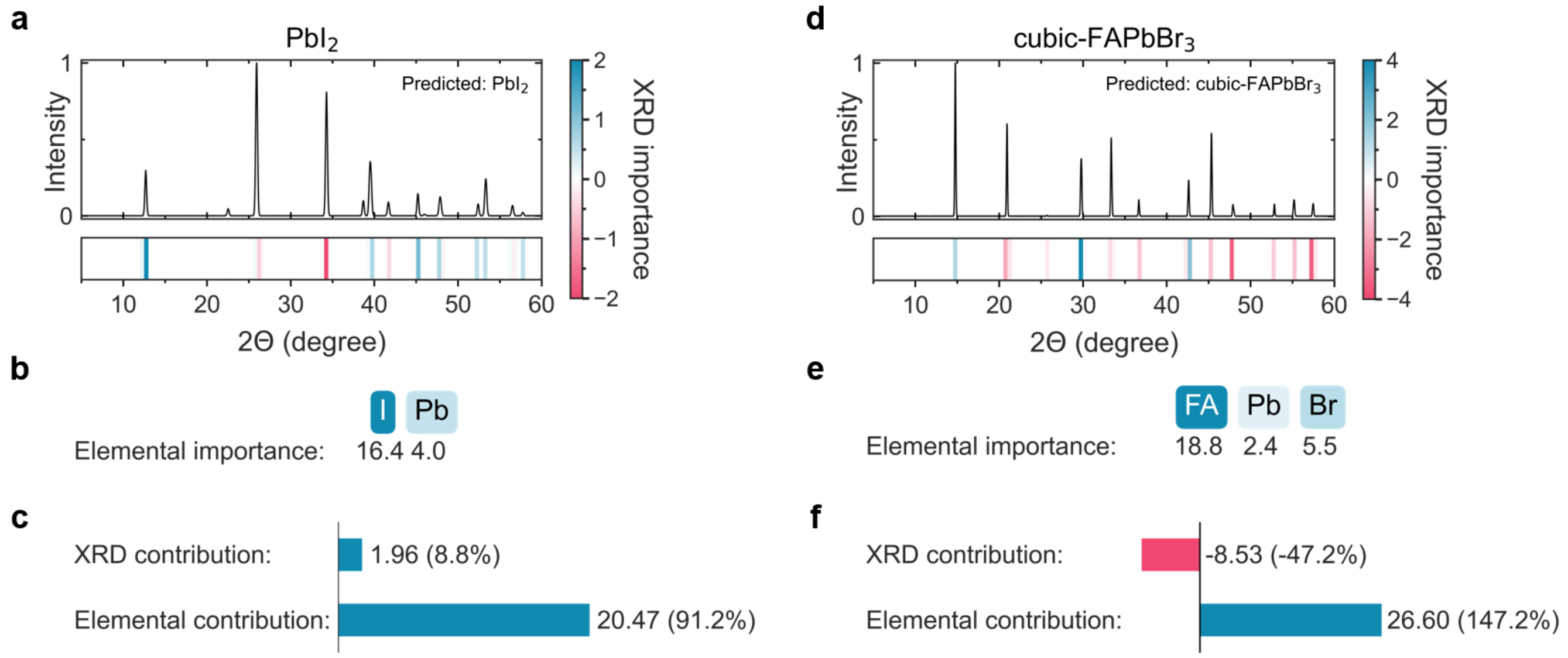

**Supplementary Figure 5. Model interpretation of elemental and structural contributions for unique phases of precursor combinations. a-c**, Individual contributions of (**a**) XRD and (**b**) elemental modalities, as well as (**c**) the overall contributions to the final ranking score for classifying $PbI_2$. **d-e**, Contributions for classifying cubic $FAPbBr_3$.

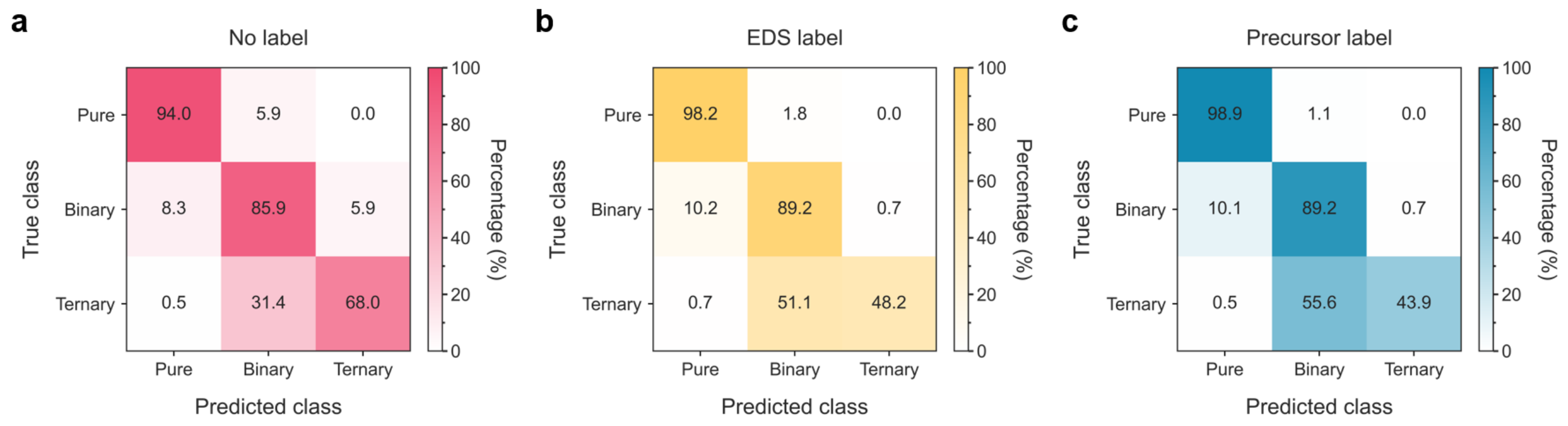

**Supplementary Figure 6. Confusion matrix for phase mixture classification of Chem-XRD.** Values show the percentage distribution of model predictions with different elemental priors: (a) no label, (b) EDS label, and (c) precursor label. Diagonal elements represent correct classifications.

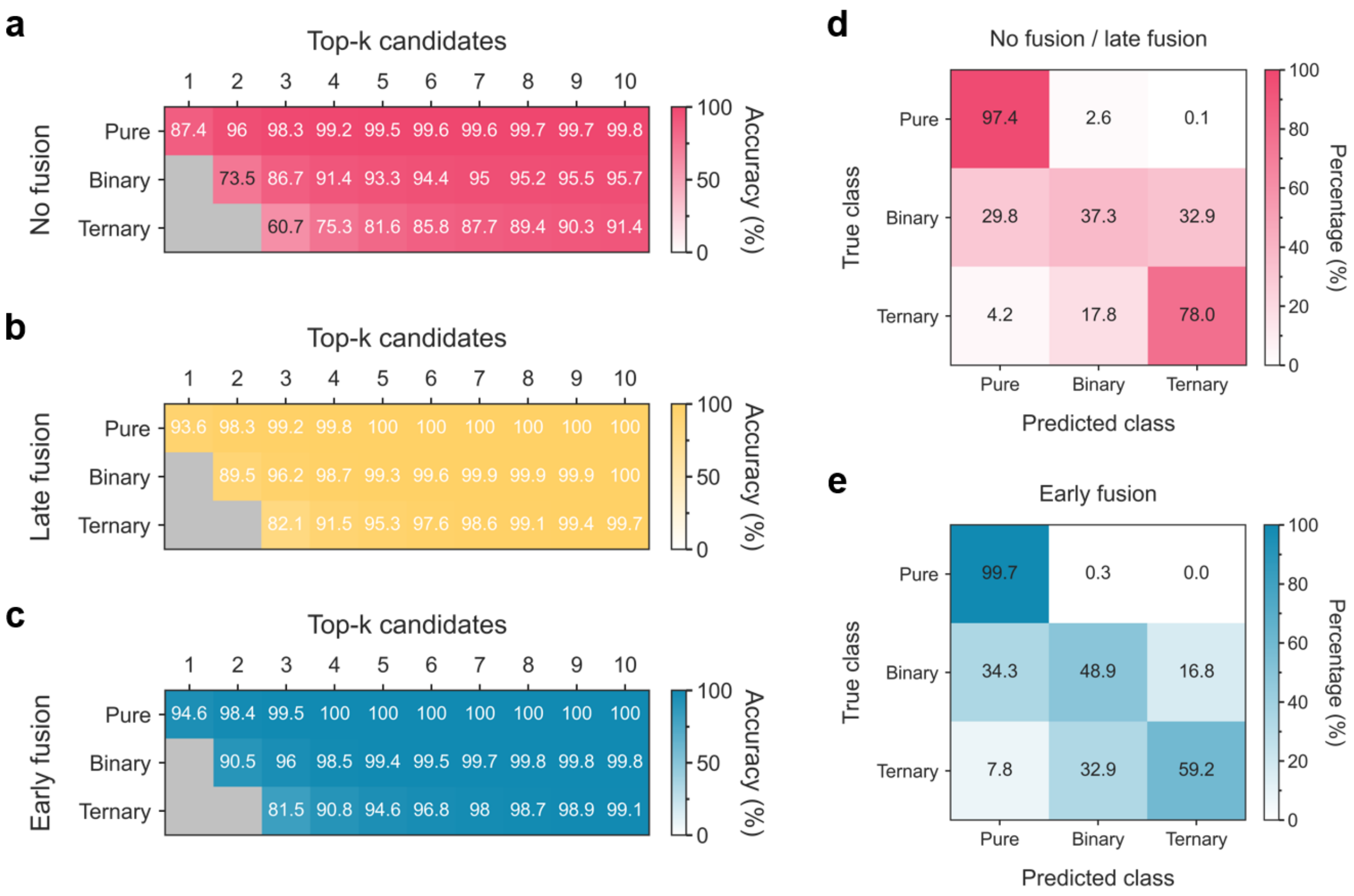

**Supplementary Figure 7. Multiphase XRD identification of convolutional neural network (CNN) models with different fusion strategies. a-c**, Top-k accuracy with (**a**) no fusion, (**b**) late fusion, and (**c**) early fusion. **d** and **e**, the corresponding confusion matrices. A total of 30 million unique samples were used for training, resulting in 1.5 million training steps under a batch size of 20. See **Supplementary Table 2** for model architectures.

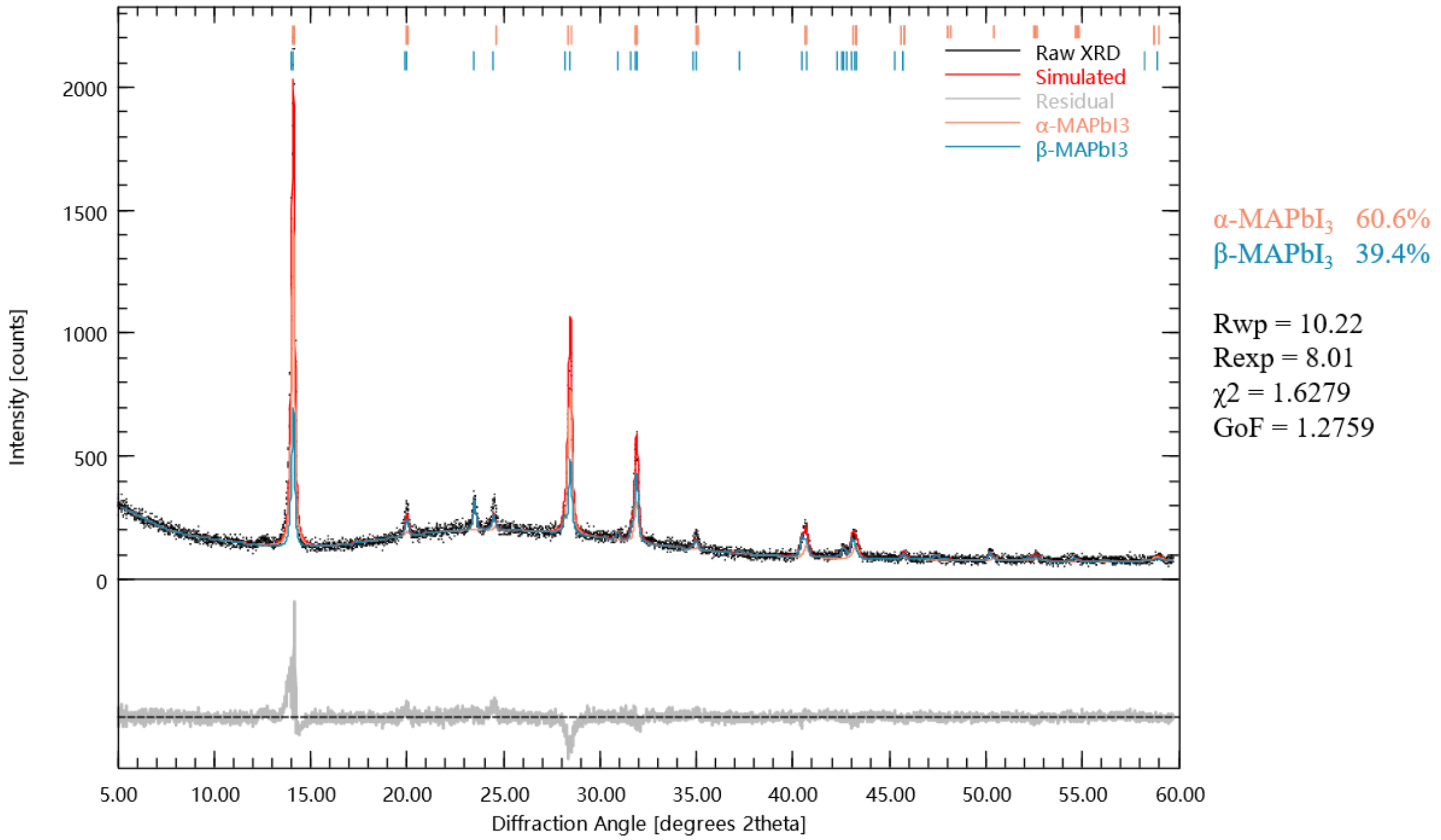

**Supplementary Figure 8. Rietveld refinement results of $MAPbI_3$ thin film XRD.** The refinement was performed using the Profex software, revealing a mixture of α-$MAPbI_3$ (60.6%) and β-$MAPbI_3$ (39.4%) phases. The residuals were calculated by subtracting the simulated XRD patterns and background signals from the absolute raw data, and were then plotted using the same y-axis spacing.

| Composition | # of ICSD entries | Down-selected material entries | | | Ref. |
|---|---|---|---|---|---|
| | | Collection code | Common name | Space group | |
| C-N-H-Pb-I | 73 | **250736** | **α-$FAPbI_3$** | **P 3 m 1** | [1] |
| | | **230492** | **β-$FAPbI_3$** | **P 4/m b m** | [2] |
| | | **230491** | **δ-$FAPbI_3$** | **P 63/m m c** | [2] |
| | | 250741 | δ-$FAPbI_3$ | P 63 m c | [1] |
| | | **250735** | **cubic $MAPbI_3$** | **P 4 m m** | [1] |
| | | **238610** | **tetragonal $MAPbI_3$** | **I 4/m c m** | [3] |
| | | 194995 | tetragonal $MAPbI_3$ | I 4 c m | [4] |
| | | 428899 | tetragonal $MAPbI_3$ | I 4/m c m | [5] |
| | | 241478 | tetragonal $MAPbI_3$ | I 4/m c m | [6] |
| | | 241484 | cubic $MAPbI_3$ | I m -3 | [6] |
| C-N-H-Pb-Br | 32 | **235795** | **cubic $FAPbBr_3$** | **P m -3 m** | [7] |
| | | 131974 | cubic $FAPbBr_3$ | P m -3 m | [8] |
| | | 131970 | orthorhombic $FAPbBr_3$ | P n m a | [8] |
| | | 131975 | tetragonal $FAPbBr_3$ | P 4/m b m | [8] |
| | | **235794** | **cubic $MAPbBr_3$** | **P m -3 m** | [7] |
| | | **33037** | **tetragonal $MAPbBr_3$** | **P 4/m m m** | [9] |
| | | 127848 | orthorhombic $MAPbBr_3$ | P m n 21 | [10] |
| | | 268784 | cubic $MAPbBr_3$ | P m -3 m | [11] |
| | | 268780 | tetragonal $MAPbBr_3$ | I 4/m c m | [11] |
| C-N-H-Pb-Cl | 5 | **243734** | **orthorhombic $MAPbCl_3$** | **P n m a** | [12] |
| Cs-Pb-I | 25 | **161481** | **α-$CsPbI_3$** | **P m -3 m** | [13] |
| | | **21955** | **γ-$CsPbI_3$** | **P b n m** | [14] |
| | | **32306** | **δ-$CsPbI_3$** | **P n m a** | [15] |
| | | 264725 | δ-$CsPbI_3$ | P n a m | [16] |
| | | 32310 | δ-$CsPbI_3$ | P n m a | [15] |
| | | 161480 | δ-$CsPbI_3$ | P n m a | [13] |
| Cs-Pb-Br | 67 | **231017** | **cubic $CsPbBr_3$** | **P m -3 m** | [17] |
| | | **84525** | **orthorhombic $CsPbBr_3$** | **P b n m** | [18] |
| | | **48997** | **$CsPb_2Br_5$** | **I 4/m c m** | [19] |
| | | **44540** | **$Cs_4PbBr_6$** | **R -3 c H** | [20] |
| | | 13938 | orthorhombic $CsPbBr_3$ | P b n m | [21] |
| | | 13939 | $CsPb_2Br_5$ | I 4/m c m | [21] |
| | | 254272 | $Cs_4PbBr_6$ | R -3 c H | [22] |
| | | 143879 | orthorhombic $CsPbBr_3$ | P n m a | [23] |
| Cs-Pb-Cl | 54 | **201251** | **cubic $CsPbCl_3$** | **P m -3 m** | [24] |
| | | **243734** | **orthorhombic $CsPbCl_3$** | **P n m a** | [12] |
| | | 143612 | orthorhombic $CsPbCl_3$ | P n m a | [25] |
| | | 23108 | cubic $CsPbCl_3$ | P m -3 m | [26] |
| | | 201250 | cubic $CsPbCl_3$ | P m -3 m | [24] |
| Cs-I | 46 | **56522** | **CsI** | **P m -3 m** | [27] |
| | | 61517 | CsI | F m -3 m | [28] |
| | | 5422 | CsI | P 4/m m m | [29] |
| Cs-Br | 7 | **236387** | **CsBr** | **P m -3 m** | [30] |
| | | 22130 | CsBr | P m n b | [31] |
| | | 22714 | CsBr | P m -3 m | [32] |
| | | 61516 | CsBr | F m -3 m | [28] |
| Pb-I | 26 | **42013** | **$PbI_2$** | **P -3 m 1** | [33] |
| | | 42510 | $PbI_2$ | R -3 m H | [34] |
| Pb-Br | 3 | **239760** | **$PbBr_2$** | **P n m a** | [35] |
| | | 36170 | $PbBr_2$ | P m n b | [36] |
| | | 202134 | $PbBr_2$ | P n a m | [37] |

**Supplementary Table 1. ICSD crystallographic information files for constructing the synthetic XRD dataset.** Material entries in **bold** meet our down-selection criteria (see also **Methods**) and were used for model training and testing. MA: methylammonium. FA: formamidinium. Note that we used a nearly cubic ICSD entry (P3m1 trigonal phase) for α-$FAPbI_3$, which was widely cited in the perovskite community, as no cubic phase $FAPbI_3$ (Pm-3m) entry exists in the ICSD database. We found no ICSD entry for β-$CsPbI_3$. We excluded CsCl and $PbCl_2$, as they are rarely reported in the literature as impurities for other halide perovskite phases.

Single-phase classification

| Fusion type | | No fusion | | | | Late fusion | | | | Early fusion | | | |
|---|---|---|---|---|---|---|---|---|---|---|---|---|---|
| Model | | CNN | | Chem-XRD | | CNN | | Chem-XRD | | CNN | | Chem-XRD | |
| Metric | | Acc % | F1 % | Acc % | F1 % | Acc % | F1 % | Acc % | F1 % | Acc % | F1 % | Acc % | F1 % |
| Data size | 100 | 1.8 | 0.8 | 1.2 | 0.0 | 28.2 | 17.9 | 15.1 | 10.1 | 3.5 | 1.5 | 4.0 | 1.5 |
| | 300 | 6.2 | 3.1 | 1.7 | 0.1 | 26.4 | 17.0 | 12.7 | 7.0 | 3.8 | 2.3 | 7.2 | 2.5 |
| | 1k | 5.5 | 3.2 | 1.7 | 0.1 | 31.1 | 20.8 | 9.6 | 6.4 | 10.3 | 9.3 | 7.4 | 3.6 |
| | 3k | 42.4 | 36.6 | 1.6 | 0.0 | 65.0 | 60.8 | 19.8 | 9.3 | 50.9 | 44.5 | 7.3 | 3.2 |
| | 10k | 76.3 | 75.3 | 3.0 | 0.2 | 86.8 | 86.4 | 32.9 | 21.8 | 74.9 | 72.8 | 34.3 | 20.8 |
| | 30k | 81.3 | 80.2 | 6.3 | 1.9 | 88.6 | 88.4 | 48.4 | 43.7 | 82.8 | 81.9 | 41.4 | 30.3 |
| | 100k | 84.4 | 83.7 | 21.8 | 18.6 | 90.8 | 91.1 | 64.8 | 63.5 | 84.2 | 83.9 | 63.7 | 60.3 |
| | 300k | 84.7 | 85.0 | 55.8 | 56.1 | 92.0 | 92.6 | 79.8 | 80.4 | 87.5 | 87.9 | 79.3 | 79.1 |
| | 1M | 85.5 | 85.7 | 68.2 | 69.1 | 92.5 | 93.3 | 84.2 | 84.9 | 93.0 | 93.7 | 87.1 | 87.3 |
| | 3M | 86.4 | 86.1 | 70.2 | 71.5 | 92.8 | 93.2 | 84.7 | 85.3 | 94.2 | 94.7 | 88.4 | 88.9 |
| | 10M | 88.7 | 89.0 | 73.8 | 74.8 | 94.0 | 94.5 | 86.4 | 87.1 | 95.2 | 95.7 | 90.3 | 91.1 |
| | 30M | 91.5 | 91.8 | 78.4 | 79.3 | 96.0 | 96.4 | 90.5 | 90.9 | 96.4 | 96.8 | 93.4 | 93.7 |

**Supplementary Table 2. Classification results for different fusion strategies on simulated single-phase XRD data.** The results are color-coded, with the best values in green and the worst in red. An underline indicates the best-performing model for each training data size. Acc: accuracy. F1: F1 score.

We used a customized CNN model which consists of two 1D convolutional layers and three fully connected layers. The first convolutional layer used 64 filters with a kernel size of 50 and a stride of 2, followed by a max-pooling layer with a kernel size of 3 and a stride of 2. The second convolutional layer used 64 filters with a kernel size of 25 and a stride of 3, followed by a max-pooling layer with a kernel size of 2 and a stride of 3. The output was flattened and passed through fully connected layers with 2000, 500, and $n$ output units, respectively, where $n$ corresponds to the total number of labels. The CNN model was trained with a batch size of 20 and a learning rate of $10^{-5}$ with an ADAM optimizer. To achieve the early fusion, the elemental information was encoded by either 1 (present) or 0 (not present) with a fixed total sequence length equal to the total amount of available element labels. The resulting encoded elemental labels were concatenated with the XRD patterns as the initial input to the model.